# Acoustic characteristics of FeSe single crystals

G.A. Zvyagina[1], T.N. Gaydamak[1], K.R. Zhekov[1], I.V. Bilich[1], V.D. Fil[1], D.A. Chareev[2], and A.N. Vasiliev[3]

[1] B.Verkin Institute for Low Temperature Physics and Engineering of the National Academy of Sciences of Ukraine, 47 Lenin Ave., 61103, Kharkov, Ukraine
[2] Institute of Experimental Mineralogy, 142432, Chernogolovka, Moscow Region, Russia
[3] Low Temperature Physics and Superconductivity Department, Moscow State University, 119991, Moscow, Russia

**Abstract** – The results of the comprehensive ultrasonic research of high quality single crystals of FeSe are presented. Absolute values of sound velocities and their temperature dependences were measured; elastic constants and Debye temperature were calculated. The elastic $C_{11}$-$C_{12}$ and $C_{11}$ constants undergo significant softening under the structural tetra-ortho transformation. The significant influence of the superconducting transition on the velocity and attenuation of sound was revealed and the value of the superconducting energy gap was estimated.

Despite considerable interest to the family of iron-based superconductors, systematic experimental data concerning the behaviour of their elastic properties are absent now. As a rule, there is some information about compressibilities that are received from structural investigations under pressure. The complete information can only be obtained by an acoustic study of single crystals. These data are useful for a verification of theoretical approaches used for the calculations of the energy spectra and for the discussion of possible technical applications in view of very high critical fields of these materials. FeSe is a key member of the binary system "11". Despite its relatively low transition temperature (~8K) it is of great interest for the research. This is due to the extreme simplicity of its structure that allows to expect close agreement of theoretical calculations to the experiment and is also due to the possibility of growing high-quality single crystals of suitable for the measurements size. FeSe exhibits one of the largest pressure effects on transition temperature [1]. It was found that FeSe has extremely low values of bulk modulus [1,2]. The investigation of the relaxation of photoexcited electron states in FeSe has revealed the features, which indicate giant softening of phonon subsystem under structural tetra-ortho transformation [3]. One order of magnitude lower softening, but still considerable one, was observed under the superconducting transition [3]. In this report we present results of the comprehensive ultrasonic research of tetragonal FeSe single crystals. Absolute values of sound velocities for the main acoustic modes and their temperature dependences have been measured. It was shown that under the structural transition $C_{11}$-$C_{12}$ as well as $C_{11}$ modes have undergone significant softening. Some features of the absorption of $C_{11}$ mode have pointed to the anomalously strong interaction of electrons with this mode, and have permitted us to estimate the value of the superconducting energy gap.

*a) Elastic constants*

The samples of FeSe$_{0.963\pm0.005}$ were grown by the same technology as in [3,4]. They have shapes of platelets with the typical size 2×2×0.4mm with [001] axis orthogonal to the plane of the platelet. The high quality of the samples was confirmed by observation of clearly defined λ-anomaly of heat capacity at superconducting transition [4]. The samples for the acoustic measurements with plane–parallel sides were prepared by grinding with fine abrasive powder (the grain size ~1μm). Measuring equipment is described in Ref. [5]. The measurements of the absolute velocity have been performed at the liquid nitrogen temperature with the use of the "nonius" procedure at frequencies ~ 55MHz. The latter consists of the measuring of the phase shift contributed by a sample $\Phi = 2\pi n + \delta\Phi$ ( $0 \le \delta\Phi \le 2\pi$ ) in two stages. First, the integer number of wavelengths $n$, which can be involved in the sample, is determined, and then, by using the correction value $\delta\Phi$, the real velocity is calculated. The method provides the accuracy of measurements better than 1% in samples of submillimeter sizes.

It is well known that in FeSe at T ~ 90K there is a structural tetra–ortho transition, but due to the formation of a domain structure the global symmetry for the bulk specimen stays tetragonal. No efforts for monodomenization were undertaken, and our results are averaged over polydomain structure. That is why we use the tetragonal indication for any temperature.

Sound velocities at T = 77K are presented in table 1.

In this table the velocity of $C_{11}$-$C_{12}$ mode is absent. Because of large attenuation of this mode below 100K we could not obtain reliable results even for the minimal length (~ 0.25mm) of the sample.

Table 1: Sound velocities in FeSe at T = 77K.

| mode | $C_{11}$ | $C_{33}$ | $C_{66}$ | $C_{44}$ | C'[a] |
|---|---|---|---|---|---|
| S ($10^5$) cm/s | 2.94 | 2.69 | 2.14 | 1.38 | 3.59 |

[a] C'=0.5($C_{11}$+$C_{12}$)+$C_{66}$.

The behaviour of elastic modules, which have been calculated by using measured temperature variations of sound velocities (we used X-ray density $\rho = 5.65 g/cm^3$), is presented in Fig.1. Note that the $C_{12}$ module was found from the velocity of the C' mode.

Technologically, we were not able to produce the sample, suitable for the measurements of the value of the $C_{13}$ mode. The latter was calculated for 50K and 190K by using the known value of the volume modulus $B$ [2].

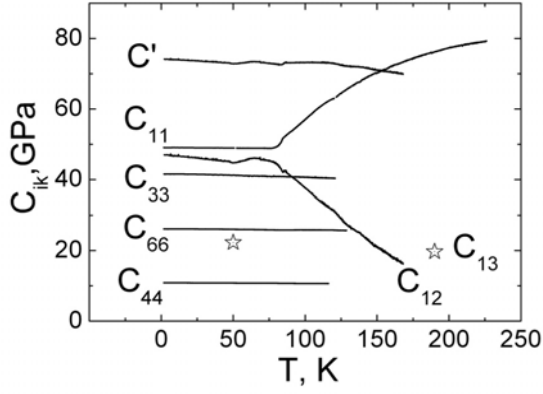

Fig.1: Temperature dependences of various components of the tensor of the elastic modules. Stars are values of $C_{13}$ which were calculated from the published data of the bulk compressibility [2].

The equation is:

$$C_{13} = B \pm \sqrt{(B - C_{33})\left(B - \frac{C_{11} + C_{12}}{2}\right)}$$

From two opportunities we need to choose the variant, which provides suitable to the experiment [2] positive signs of the partial compressibilities along the main axes (in our case the lower sign was chosen). We can check the consistency of our measurements with the data of structural investigation under pressure [2]. Let us compare the ratio of the partial compressibilities along the $c$ axis and in the $ab$ plane. In view of the aforementioned averaging over the domain structure we have:

$$\frac{2K_c}{K_a + K_b} = \frac{C_{11} + C_{12} - 2C_{13}}{C_{33} - C_{13}}$$

From our data we get $2.6 \pm 0.1$ (50K) and $2.5 \pm 0.1$ (190K), whereas from Ref. [2] we have 2.67 and 2.57 respectively.

Let us turn attention to the unconventional behaviour of elastic constants under the tetra-ortho structural transition. As a rule there is only one soft mode ($C_{66}$, or $C_{11}-C_{12}$), which experiences a strong softening near the transition point, but it rapidly recovers it's stiffness under further lowering of the temperature. In our case two modes ($C_{11}$ and $C_{11}-C_{12}$) do not just experience a considerable softening near the transition point, but they still stay «soft» under further lowering of the temperature. A similar behaviour of a longitudinal acoustic mode was also found in Ref. [3].

In Fig.2 the attenuation of the $C_{11}$ mode is shown. It demonstrates an abnormally large increase below the temperature of the phase transition. As a rule, the attenuation of the mode active under phase transition has a sharp maximum near the critical temperature that is caused by scattering of sound on order-parameter fluctuations. In our case the excess attenuation is observed in the whole low temperature interval. One might think that it is the consequence of the scattering of this mode on the polydomain structure with significant mismatch of the acoustic impedances of individual domains.

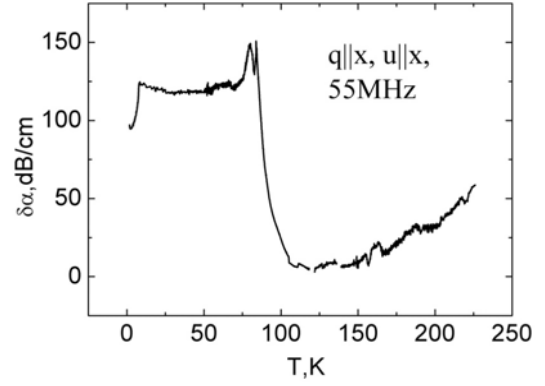

Fig. 2: Temperature dependence of the changes of the attenuation of the $C_{11}$ mode.

Another observed feature is the unusual behaviour of the sound velocities at low temperatures: All studied modes except of the $C_{66}$ below the structural transition temperature show a linear change (Fig.3) ($C_{66}$ mode varies as $T^2$). As a rule, such behaviour takes place in amorphous and disordered systems [6,7].

The Debye temperature was calculated using the formula (in CGS units) [8]:

$$\Theta_D = 0.003625\left(\frac{\rho n}{MI}\right)^{\frac{1}{3}}$$

Here $n = 2$ is the number of sites for a molecule, $M$ is the molecular weight, $I$ is the sum of reverse cubes of sound values averaged over directions of the wave normal. For the orthorhombic phase we have $\Theta_{D\ ortho} = 159K$. This value is slightly lower than in Ref [4] (210K). For the tetragonal phase temperature is $\Theta_{D\ tetra} \approx 200K$.

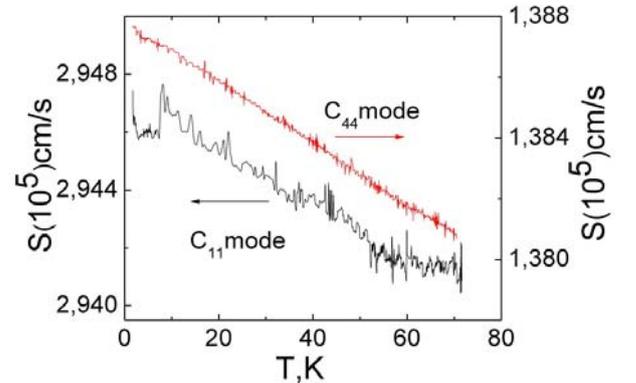

Fig. 3: Low temperature dependences of the sound velocities.

Elastic constants for the tetragonal $FeSe_{1-x}Te_x$ system, including x = 0, were calculated in [9]. Only the longitudinal constants ($C_{11}$ and $C_{33}$) are close to the experimental velocities; for the rest the correlation is practically absent.

*b) Influence of superconducting transition*

The phenomenological theory of type II phase transitions gives the expression for the changes of the elastic constants in the superconducting state (see, for example, [10]):

$$\delta C = -(\frac{dT_C}{d\varepsilon})^2 \frac{\Delta C_Q}{T_C} + A_1 \frac{T}{T_C}(1 - \frac{T}{T_C}) + A_2 (1 - \frac{T}{T_C})^2 \quad (1)$$

Here $\varepsilon$ is the strain corresponding to modulus $C$, $\Delta C_Q$ is the jump of the heat capacity per unit volume at $T_C$, constants $A_1$ and $A_2$ are formed from the first- and second-order derivatives of $T_C$ and the condensation energy with respect to the deformation [10]. The first term refers only to the longitudinal waves and represents step-like lowering of the related module in $T_C$. The second and third terms describe the evolution of the module below $T_C$. In Fig.4 the changes of the velocities of the longitudinal modes below the temperature of superconductor transition are given. In fact, these changes are jumps imposed on the background linear dependence.

They are abnormally large and exceed top values observed for heavy fermions systems [10], and it is the result of a very high pressure dependence of $T_C$. One can see from Fig. 4 that the effect is practically isotropic. Therefore, we can use the value $dT_C/dp$ under the hydrostatic pressure, and the value of the bulk module as the elastic one in our estimates. Then the expression for the jump of the sound velocity can be written as:

$$\frac{\delta S}{S} = \frac{B}{2}(\frac{dT_C}{dp})^2 \frac{\Delta C_Q}{T_c}$$

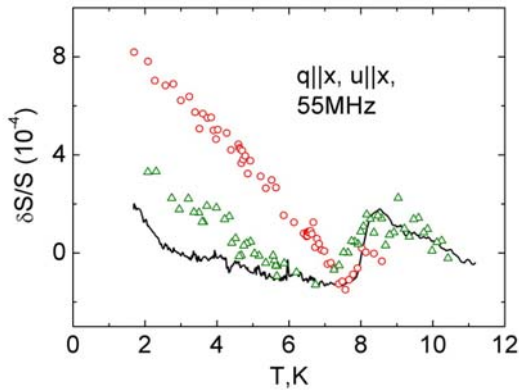

Fig. 4: The influence of the superconducting transition on the velocity of the longitudinal sound waves. Full line corresponds to the $C_{11}$ mode, circles - to the C' mode, triangles - to the $C_{33}$ mode.

Using $B = 33$ GPa [2], $dT_C/dP = 7$ K/GPa [1] and $\Delta C_Q/T_C = 3.4 \cdot 10^3$ erg/cm$^3 \cdot$K$^2$ [4], we get $\delta S/S \sim 2.5 \cdot 10^{-4}$, which is in good agreement with Fig. 4. Two-orders-of-magnitude larger softening of the energy of the longitudinal phonon mode in $T_C$ was reported in [3]. Perhaps this contradiction is due to the fact that the experiments [3] had deal with much higher phonon frequencies ($\omega \sim 10^{11}$ s$^{-1}$).

In Fig.5 the variation of the velocity and attenuation of the transverse $C_{66}$ mode is shown.

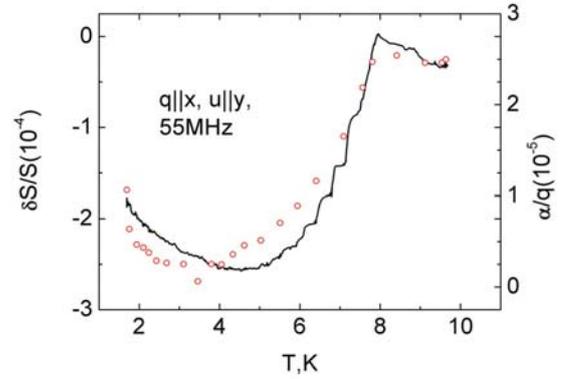

Fig. 5: The influence of the superconducting transition on the velocity and the attenuation of the $C_{66}$ mode. Full line corresponds to the velocity, circles – to the attenuation.

The velocity is well described by the second term in the equation (1) with $A_1 \approx -0.001$. Unlike the longitudinal waves the effect is anisotropic: $C_{44}$ mode does not feel any significant changes in $T_C$.

Already from Fig.2 it is seen that the attenuation of the $C_{11}$ mode experiences significant changes at $T_C$.

Apparently, we deal with the unique situation for the modern superconducting systems: A measurable interaction of ultrasound with electrons. It is well known that the structure of the absorption coefficient of the sound for the Fermi surface (FS) of the general view has the form:

$$\alpha/q = (\Lambda/\varepsilon_F)^2 \cdot S/v_F \cdot f(ql)$$

Here $\Lambda$ is the deformation potential, $\varepsilon_F$ is the Fermi energy, $v_F$ is the Fermi velocity, $q$ is the wave number, $l$ is the mean free path. The last multiplier describes the frequency dependence.

For $ql < 1$ $f(ql) \sim ql$, and the absorption is quadratic in frequency. In the opposite case $f(ql)$ and $\alpha/q$ are constants. If the FS has a large flat section, and the electrons on it interact effectively with the sound, the multiplier $S/v_F$ must be changed to the relative area of the flattening [11]. The results of a more detailed study for the attenuation of different modes below $T_C$ are shown in Fig.6 (as the referent point we used value of the signal at T = 2.5K), from which we can make some unexpected conclusions:

1) The absorption of the $C_{11}$ mode at the wavelength has a record value for metals, including high purity ones [12]. This may be due to the large value of $\Lambda$, or due to flattening of the FS.

2) The absorption coefficient varies with frequency almost linearly. In any case the fulfillment of the condition $ql \geq 1$ is required.

3) The effect is very anisotropic. For $C_{33}$ and C' modes it is not observed.

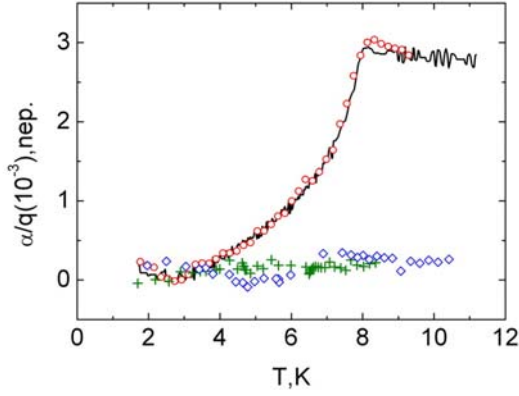

Fig. 6: The influence of the superconducting transition on the attenuation of the longitudinal sound. Full line corresponds to the $C_{11}$ mode. (F = 55MHz), circles – to the $C_{11}$ mode (100MHz), crosses – to the C' mode (55MHz), diamonds – to the $C_{33}$ mode (55MHz).

Let us estimate the relaxation time τ. The specific resistance of the samples, analogous to ours, is near 40 μOhm·cm [3]. The density of carriers from the measurements of the Hall effect is about $10^{20} \div 10^{21} cm^{-3}$ [13]. Since the electronic structure of FeSe is close to the one of the compensated metal [14], these values are rather overestimated. It gives us $\tau \sim 10^{-12}$s and $ql \sim 10^{-1}$, i.e. our samples are not too far from a pure limit.

However, it is improbable to expect that the parameter $ql$ is close to unity on the whole FS. In this case it would be expected that the magnetic-field dependence of the attenuation was the manifestation of the magneto-acoustic geometric and quantum oscillations, accompanied by a noticeable change in the monotonic part. All our attempts to detect any changes of a signal with the strength of the magnetic field in fields up to 5T were unsuccessful. Perhaps FS has a preferred direction in which the mean free path has a sharp maximum, which leads to the observed characteristics of the absorption.

Bardeen-Cooper-Schrieffer theory of superconductivity irrespective to the value of the parameter $ql$ gives the dependence for the relation between the attenuation in superconducting state and the one in a normal state:

$$\frac{\alpha_s}{\alpha_n} = \frac{2}{\exp(\frac{\Delta(T)}{kT}) + 1}$$

Here $\Delta(T)$ is the energy gap. This equation was deduced for an isotropic superconductor. Actually, it is valid in any nodeless model of superconductivity. In anisotropic superconductors (or multiple-gap ones) the Δ(T) should be understood as some effective quantity approximately equal to the minimum value of the energy gap for the electrons that interact with the sound mode (see [12] and references therein). So it is possible to estimate the gap value by inverting this equation, (Fig. 7).

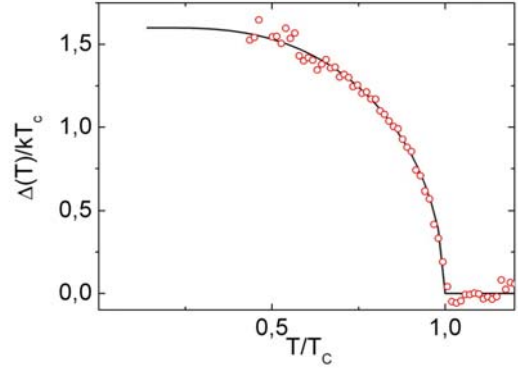

Fig. 7: Temperature dependence of the value of the energy gap. Circles correspond to the experiment, full line corresponds to the BCS dependence for $\Delta(0) = 1,6kT_C$.

It yields somewhat lower value than the BCS one, however it practically coincides with the data from [3] (~ 1.62 $kT_C$). The temperature dependence of the gap is identical to the theoretical one.

In summary, we have determined the components of the tensor of elastic constants for FeSe single crystal. The structural phase transition is accompanied by the significant softening of the $C_{11}$ and $C_{11} - C_{12}$ modes. We have observed a significant influence of the superconducting transition on the velocity and attenuation of sound and estimated the value of the superconducting energy gap as 1.6 $kT_C$.

These investigations were partly supported by the grants of President of Russian Federation MK-1557.2011.5, the Ministry of Education and Science of Russian Federation 83-78 and 11.519.11.6012, and Russian Foundation for Basic Research 12-02-90405.

REFERENCES

[1] MARGADONNA S., TAKABAYASHI Y., OHISHI Y., MIZUGUCHI Y., TAKANO Y., KAGAYAMA T., NAKAGAWA T., TAKATA M., and PRASSIDES K., *Phys. Rev.,* **B80**, (2009), 064506.
[2] MILLICAN J. N., PHELAN D., THOMAS E. L., LEÃO J. B., and CARPENTER E., *Solid State Commun.*, **149**, (2009), 707.
[3] LUO C.W., WU I. H., CHENG P. C., LIN J.-Y., WU K. H., UEN T. M., JUANG J.Y., KOBAYASHI T., CHAREEV D. A., VOLKOVA O. S., and VASILIEV A. N., *Phys. Rev. Lett.* **108**, (2012), 257006.
[4] LIN J.-Y., HSIEH Y. S., CHAREEV D. A., VASILIEV A. N., PARSONS Y., and YANG H. D., *Phys. Rev.* **B84**, (2011), 220507.
[5] MASALITIN E. A., FIL V. D., ZHEKOV K. R., ZHOLOBENKO A. N. and IGNATOVA T. V., *Low Temp. Phys.,* **29,** (2003), 72.
[6] BELESSA G., *Phys.Rev .Lett.* **40** (1978), 1456.
[7] ZHERLITSYN S. V., POPOVA V. V., FIL V. D., and KHABIBULLIN KH. G., *Fizika Nizkikh Temp.* **15**, (1989), 24.
[8] FEDOROV F.I., *Theory of Elastic Waves in Crystal,* (Plenum Press, New York) (1968).
[9] CHANDRA S. and ISLAM A. K. M. A., *Physica (Amsterdam),* **470C**, (2010), 2072.
[10]   LÜTHI B., ZHERLITSYN S., and WOLF B, *Eur. Phys. J.***, B46**, (2005), 169.

[11] KONTOROVICH V.M.: *Sov. Phys. Usp.* **27**, (1984), 134.
[12] RAYNE J.A. and JONES C.K., *Physical Acoustics*, **Vol. VII,** (Acad. Press, 1970), 149.
[13] FENG Q. J., SHEN D. Z, ZHANG J. Y, LI B. S,. LI B. H, LU Y. M., FAN X. W., and LIANG H. W., *Appl. Phys. Lett.,* **88**, (2006), 012505.
[14] MA F., JI W., HU J., LU Z.-Y., and XIANG T., *Phys. Rev. Lett.*, **102**, (2009), 177003.